\documentclass[twocolumn,showpacs,preprintnumbers,amsmath,amssymb,A4paper,prl]{revtex4}
\usepackage{graphicx}% Include figure files
\usepackage{dcolumn}% Align table columns on decimal point
\usepackage{bm}% bold math

\begin{document}
\newcommand{\kp}{{\bf k$\cdot$p}\ }

\preprint{APS/123-QED}

\title{One-dimensional semirelativity for electrons in carbon nanotubes}
%\\with Forced Linebreak}% Force line breaks with \\

\author{Wlodek Zawadzki}
 \affiliation{Institute of Physics, Polish Academy of Sciences\\
       Al. Lotnikow 32/46, 02--668 Warsaw, Poland\\
 %  e-mail: zawad@ifpan.edu.pl
 }

\date{\today}% It is always \today, today,

\begin{abstract}
It is shown that the band structure of single-wall semiconducting carbon nanotubes (CNT) is analogous to
relativistic description of electrons in vacuum, with the maximum velocity $u$= $10^8$cm/s replacing the light
velocity. One-dimensional semirelativistic kinematics and dynamics of electrons in CNT is formulated. Two-band
\textbf{k}$\cdot$\textbf{p} Hamiltonian is employed to demonstrate that electrons in CNT experience a
Zitterbewegung (trembling motion) in absence of external fields. This Zitterbewegung should be observable much
more easily in CNT than its analogue for free relativistic electrons in vacuum.

\end{abstract}

%PACS numbers: {73.22.-f  73.63.Fg   78.67.Ch} %
\maketitle Since the first reported observations of carbon nanotubes [1,2] these unique one-dimensional
nanostructures were subject of very intensive research, both because of their remarkable properties as well as
their potential use in nanometer-sized electronics (see Refs. 3 and 4). Among other particularities CNT have
interesting energy band structure and it is this aspect that is of our concern here. We will be interested in
the simplest single-wall semiconducting CNT. Such tubes are obtained from a slice of graphene wrapped into a
seamless cylinder, so that the 1D band structure of CNT can be constructed by using the 2D band structure of
graphene. The periodic boundary condition around the tube circumference causes quantization of the transverse
wave vector component $k_x$. The purpose of our work is to predict new properties of electrons and holes in CNT
in the classical and quantum domains. To this end we use a similarity of the band structure of CNT to the
relativistic description of free electrons in vacuum. In particular, we predict that the "semirelativistic" band
structure of CNT should result in Zitterbewegung (trembling motion) of electrons in absence of external fields.
Similar phenomenon was predicted for relativistic electrons in vacuum but never observed. Thus, an observation
of this effect in CNT would be of great importance not only for the solid state physics but also for the high
energy physics.

  In the following we use the\textbf{ k$\cdot $p} band structure at the K point of the Brillouin zone [5,6].
  The initial $2\times2$
  Hamiltonian is written in the form
\begin{equation}
\hat{H}=\alpha\left[\begin{array}{cc} 0 & a_n-i\hat{p}\\a_n+i\hat{p} & 0\end{array} \right]\;\;,
\end{equation}
where $\alpha$ is a coefficient, ${\hat p}$ is the operator of pseudomomentum in the $y$ direction and $a_n$ is
given by the quantization of the wave vector $k_x$. There is $a_n=\hbar k_x(n)=\hbar(2\pi/L)(n-\nu/3)$ for $n=0,
\pm1, \pm2...$. Here $L$ is the length of circumference. Semiconducting CNT of our interest are obtained for
$\nu=\pm1$. In absence of external fields the resulting energy is ${\cal E} =\pm E(p)$, where
$E(p)={\alpha}(a_n^2+p^2)^{1/2}$. The upper sign is for the conduction and the lower for the valence band. The
above relation is analogous to the dispersion $E(p)$ for free relativistic electrons in vacuum. We write the
energy in the following equivalent form
\begin{equation}
E(p)=\left[\left(\frac{{\epsilon}_g}{2}\right)^2 + {\epsilon}_g \frac{p^2}{2m^*_0}\right]^{1/2}\;\;,
\end{equation}
 where
$\epsilon_g = 2\alpha a_n$ is the energy gap and $m_0^* = a_n/\alpha$ is the effective mass at the band edge,
related to band's curvature for small $p$ values. Both $\epsilon_g$ and $m_0^*$ have different values for
different subbands $n$. Equation (2) has the relativistic form with the correspondence: $\epsilon_g \rightarrow
2m_0^*c^2$ and $m_0^* \rightarrow m_0$ [7].

 We
first investigate some consequences of the one-dimensional dispersion (2). The electron velocity is
$v=dE/dp=\alpha^2p/E$. For large momenta the velocity reaches a saturation value
$u=\alpha=(\epsilon_g/2m^*_0)^{1/2}$, the same for all subbands. The maximum velocity $u$ plays for electrons in
CNT the role of the light velocity $c$ for relativistic electrons in vacuum. Using $u$ we can write the energy
in another equivalent form
\begin{equation}
E(p)=[(m^*_0u^2)^2+u^2p^2]^{1/2},
\end{equation}
 which is directly reminiscent of the relativistic $E(p)$ relation.
Now we define an energy-dependent effective mass $m^*$ relating velocity to momentum
\begin{equation}
m^*v=p.
\end{equation}
We calculate $m^*=p/v=E/u^2.$ This gives
\begin{equation}
E=m^*u^2,
\end {equation}
 which is in one-to-one correspondence to the Einstein relation $E=mc^2$ (the maximum
velocity $u$ replacing $c$). For $p=0$ Eqs. (3) and (5) reduce to $E_r=m_0^*u^2$, which corresponds to the
formula for the rest energy. One can also express the mass $m^*$ by the velocity. Beginning with the relation
$p^2u^2=v^2E^2/u^2$ , employing $E^2=(m_0u^2)^2+u^2p^2$ and solving for momentum, we get
\begin{equation}
p=m_0^* \gamma v,
\end{equation}
 where $\gamma=(1-v^2/u^2)^{-1/2}$. Using the above definition $p=m^*v$ we have $m=m_0^*\gamma$, which has the
familiar relativistic form (with $u$ replacing $c$). Now Eq.(5) reads $E=m_0^*\gamma u^2$.

 Next we assume, in
analogy with the special relativity, that $dp/dt=F$, where $F$ is the force. One can now define an another
effective mass $M^*$, relating force to acceleration
\begin{equation}
M^*a=F.
\end{equation}
  Since $a=dv/dt=(dv/dp)(dp/dt)=(d^2E/dp^2)F$, we obtain $1/M^*=d^2E/dp^2$. Using the above
dispersion $E(p)$ we get
\begin{equation}
M^*=\frac{E^3}{m^{*2}_0u^6}=m_0^*\gamma^3,
\end{equation}
 which again has the corresponding relation in special relativity if the acceleration is parallel to the force.

   Estimating the introduced quantities, we use the value of $\alpha \hbar=6.46$ eV$\AA$ [6] . This
 gives for the maximum velocity $u=\alpha=0.98\times10^8$cm/s. The lowest energy gap is $\epsilon_g(0)=2\alpha a_0$,
where $a_0=\hbar 2\pi/3L$. For the circumference $L=60\AA$ we get $\epsilon_g(0)=0.45$eV. The effective mass
$m^*_0/m_0=a_0/\alpha m_0=0.041$ for the same conditions. The quoted parameters are close to those of the the
typical narrow gap semiconductor InAs, but CNT of higher diameter have smaller $\epsilon_g$ and $m_0^*$. We
emphasize that, while for a nonparabolic energy band one can legitimately define both energy
   dependent masses $m^*$ and $M^*$, it is the "momentum" mass $m^*$ that is a more useful quantity. This mass is employed
   in the transport theory since the current is related to velocity, not to acceleration. It is $m^*$ that enters
   into the equivalence of the mass and the energy in Eq.(5). Finally, $m^*$ is related to the density of electron
   states since the latter is $\rho(E)=(1/\pi \hbar)(dp/dE)=(1/\pi \hbar)(m^*/p)$.

 Next we consider the quantum effects in CNT related to the Hamiltonian (1). To this purpose we introduce an
 important quantity [8]
 \begin{equation}
 \lambda_Z=\frac{\hbar}{m_0^*u}=\frac{\hbar}{a_n},
 \end{equation}
 which we call the length of Zitterbewegung (see below). It corresponds to the Compton wavelength $\lambda_c$ for
 electrons in vacuum and it plays for the semirelativistic band structure (1) the role that $\lambda_c$ does for the Dirac
 equation. However, $\lambda_Z$  is several orders of magnitude larger than $\lambda_c$.
 Using $m_0^*=0.041m_0$ and $u=0.98\times10^8$ cm/s we
 calculate $\lambda_Z=28.6\AA$.

 Now let us consider the operator of electron velocity $\hat{v}=d\hat{H}/d\hat{p}$. A simple calculation shows that the eigenvalues of $\hat{v}$ are,
  paradoxically, $\pm u$. This differs drastically from the classical velocity calculated above. To clear the paradox we calculate $\hat{v}(t)$.
   Using Eq. (1) we get $\hat{v}\hat{H}+\hat{H}\hat{v}=2\alpha^2\hat{p}=2u^2\hat{p}$. Hence the time
 derivative of $\hat{v}$ is
 \begin{equation}
 \frac{d\hat{v}}{dt}=\frac{i}{\hbar}2u^2\hat{p}-\frac{2i}{\hbar}\hat{v}\hat{H}.
\end{equation}
 This represents a simple differential equation for $\hat{v}$. Its solution is
\begin{equation}
\hat{v}(t)=\frac{u^2}{\hat{H}}\hat{p}+(\hat{v} _0-\frac{u^2\hat{p}}{\hat{H}})exp(-2i\frac{\hat{H}t}{\hbar}),
 \end{equation}
 where $1/\hat{H}=E_p^{-2}\hat{H}$. Thus the quantum velocity differs from the classical one by the term  that oscillates in time. Equation (11) can
 be integrated with respect to time to give the position operator $\hat{y}$ in the
 Heisenberg picture
\begin{equation}
\hat{y}(t)=\hat{y}(0)+\frac{u^2\hat{p}}{\hat{H}}t+\frac{i\hbar
u}{2\hat{H}}\hat{A}_0[exp(\frac{-2i\hat{H}t}{\hbar})-1],
 \end{equation}
  where $\hat{A_0}=(\hat{v}_0/u)-(u\hat{p}/\hat{H})
  $. The first two terms of Eq. (12) represent the classical electron motion. The third term
 describes time
 dependent oscillations with a frequency $\omega_Z=\epsilon_g/\hbar$. Since $\hat{A_0} \approx 1$ the amplitude of oscillations is
 $2\hbar u/2\hat{H} \approx \hbar/m_0^*u=\lambda_Z.$ In the relativistic quantum mechanics (RQM) of free electrons the analogous oscillations were devised by
  Schroedinger [9,10], who called them Zitterbewegung (ZB). This explains the name given above to $\lambda_Z$. We note that  the phenomenon of
   ZB goes beyond Newton's first law since we have a nonconstant velocity without a force. In RQM it is demonstrated that the ZB is
a result of interference between states of positive and negative electron energies [11]. The ZB described above
for CNT has the frequency of about $10^7$ times lower and the amplitude of about $10^4-10^5$ times higher than
the corresponding values for relativistic electrons in vacuum. We come back to the problem of ZB below.

 In order to investigate the case of a definite sign of energy we complete the Hamiltonian (1) by the potential
 $V(y)$  on the diagonal and introduce magnetic field by adding the vector potential to the momentum operator. We
 then try to separate the 2$\times$2 eigenenergy problem with the Hamiltonian (1) into two independent problems for each band. This can be easily
 done in the semiclassical approximation neglecting the noncommutativity of $\hat{p}$ with $V$. One obtains then, as
 expected, the one-band effective mass approximations for the conduction and valence bands
\begin{equation}
\hat{H}=\pm\frac{\epsilon_g}{2}\pm\frac{\hat{P}^2}{2m_0^*}+V,
\end{equation}
 where $\hat{P}=\hat{p}_y+eA_y$ is the canonical momentum.

 In absence of external potentials the two-component wave functions resulting from the Hamiltonian (1) can be
 transformed exactly into one-component functions for positive (or negative) electron energies. This is achieved
 by applying the following unitary transformation $\hat{S}$
\begin{equation}
\hat{S}=\frac{1}{\sqrt{2}}(1+\frac{\hat{\beta}\hat{H}}{E_p}),
 \end{equation}
where $\hat{\beta}=  \begin{pmatrix}1&0 \\ 0 &-1
\end{pmatrix}$. It can be  verified that $\hat{S}\hat{S}^+=1$
 and that the transformed Hamiltonian $\hat{H}_\phi$ is
\begin{equation}
\hat{H}_\phi=\hat{S}\Hat{H}\hat{S}^+=E_p\hat{\beta}.
\end{equation}
 Thus the positive and negative energies are separated. One can transform any wave function $\psi(y')$  (not only the
 eigenfunctions of the Hamiltonian (1)) from the original two-component representation to the one-component
 $\phi$ representation by the following integral transformation (cf. Refs. 12,13,14)
 \begin{equation}
 \psi'_{\pm}(y)=\int{K_{\pm}}(y,y')\psi(y')dy' \;,
 \end{equation}
 where
 \begin{equation}
 K_{\pm}(y,y')=\frac{1}{8\pi}(1 \pm\hat{\beta})\int(1+\frac{\hat{H}}{E_p})e^{ip'(y-y')}dp'.
 \end {equation}
 The subscripts $\pm$  correspond to the one-component functions related to positive or negative energy,
 respectively. The factors $(1\pm\hat{\beta})$ guarantee this property. The kernels $K_{\pm}(y,y')$  are not point transformations. Suppose we
 are interested in the eigenfunction of the electron position $\hat{y}$. It is convenient to take in the initial
 representation the unit vector $\psi(y')={1\choose0}\delta(y'-y_0)$. It then follows from Eq. (16) that the transformed one-component functions are $K_{\pm}(y-y_0)$.
The integrals of $K_{\pm}(y-y_0)$  over $Y=y-y_0$ are unity. To get an idea of the extension of $K_{\pm}(y-y_0)$
we calculate their second moment. After some manipulation we obtain
\begin{equation}
\int {Y^2K_{\pm}(Y)}dY=\frac{1}{2}\lambda^2_Z.
\end{equation}
 Thus the extension of the transformed eigenfunctions of position is  $|y-y_0|=\lambda_Z/\sqrt{2}$. In the transformed states there is
 no ZB since the transformation (14) [ or (16) ] eliminates the negative (positive) energy components of the wave
 functions.
 Following the interpretation established in RQM we are confronted with the choice between a point-like electron
 described by a two-component function (four components including spin) which experiences the ZB with the
 amplitude of $\lambda_Z $, and an electron described by a one-component function (two components including spin) for
 either positive (or negative) electron energy which does not experience the ZB but is "smeared" in the
 $y$ direction to an object of the size $\lambda_Z$. One can say that in the one-component representation the trembling
 motion is averaged into smearing. As follows from the above estimation of $\lambda_Z$, the amplitude of ZB or,
 alternatively, the smearing of electrons in CNT is quite large and it should be observable directly or
 indirectly.

 In a recent paper [15] Zitterbewegung-type oscillations were proposed using the Hamiltonian of spin splitting due
 to structure inversion asymmetry (SIA) or bulk inversion asymmetry (BIA) of the system [16,17,18]. The Hamiltonian for
 BIA has the form $\hat{H}_D=(\beta/{\hbar})(\sigma_y\hat{p}_y-\sigma_x\hat{p}_x)$, where $\sigma_x$ and
 $\sigma_y$ are the Pauli spin matrices.
 Interestingly, the Hamiltonian (1) we use for CNT has the identical form, with the minus sign in $\hat{H}_D$  replaced
 by the plus sign in Eq. (1). This similarity
 is formal because the Hamiltonian (1) is not related to spin. The description of Ref. [15] begins with a 2D case but
 finally  treats a
 quantum wire, which is a system almost identical with CNT. Considering the motion of a wave packet it shows
 that the ZB occurs in the direction perpendicular to the packet's group velocity.
 Its frequency is $\omega=\Delta E/\hbar$, where
 $\Delta E$
 is the energy splitting, and its amplitude is inversely proportional to the packet's wave vector. Our Eq. (12)
 describes the same result for the frequency, namely $\omega_Z=\epsilon_g/\hbar$. As to the amplitude, our
 result is formally also the
 same since the ZB in the $y$ direction has the amplitude $\lambda_Z=\hbar/a_n=1/k_x(n)$, where $k_x(n)$ is the
 wave vector in the $x$ direction.
 However, the latter similarity is only formal because the quantized value of $k_x$ corresponds to the standing
 wave, that is to both $+k_x(n)$ and $-k_x(n)$. Thus one should rather have the packet propagating in the free
 direction $y$
 with the wave vector $k_{y0}$. The ZB is then parallel to the $x$ direction, i.e. it occurs on the tubes's
 circumference, similarly to the wire's finite width considered in [15].

 As compared to the oscillatory electron motion proposed in [15], the ZB in CNT described above has two
 advantages. First, in the configuration discussed in Ref. [15] the growth direction creating the initial 2D system
 is parallel to $z$. It is known that, for the Bychkov-Rashba Hamiltonian related to SIA, the electron spin is
 quantized in the plane transverse to the growth direction, i.e. in the $x-y$ plane (see Ref. [18]). However, the
 calculation [15] requires that the spin be directed along the $z$ direction which contradicts the above
 quantization condition. In our case there is no requirement concerning the spin direction since the mechanism of
 ZB is not related to spin. The second advantage is more fundamental. The ZB proposed above for electrons in CNT
 is the "true" Zitterbewegung in a sense that it corresponds to the ZB for free relativistic electrons in
 vacuum, while the oscillatory motion proposed in [15] has no such correspondence. Thus the observation of ZB in
 CNT would be of great value also for the relativistic quantum mechanics. There exist in the literature
 contradictory statements concerning the observability of ZB in vacuum (see e.g. Refs. 10,19,20 ), which makes its
 analogue in solids and molecules even more interesting. It appears that the ZB in CNT can be observed
 experimentally using the high-resolution scanning-probe microscopy [21,22].

 In summary, using the analogy between the band structure of single-wall semiconducting carbon nanotubes and the description of
 relativistic electrons in vacuum we formulate the one-dimensional semirelativistic kinematics and dynamics for
 charge carriers in CNT. We also consider the quantum domain demonstrating that electrons in CNT experience the
 Zitterbewegung (trembling motion) even in absence of external fields. If the electrons are described by an
 effective one-band Hamiltonian for positive (or negative) energy, there is no ZB but the electrons should be
 treated as objects extended along the direction of CNT axis. It is emphasized that the analogous effects have
 been predicted for relativistic electrons in vacuum but they are much more difficult to observe than in CNT.

 I am pleased to thank Dr. T.M. Rusin and Dr. P. Pfeffer  for elucidating discussions. This work was supported
in part by The Polish Ministry of Sciences, Grant No PBZ-MIN-008/PO3/2003.

\end{document}